\documentclass{svmult}
\usepackage{epsfig,graphicx} 
\usepackage[bottom]{footmisc}
\usepackage{natbib,url}      
\bibpunct{(}{)}{;}{a}{}{,}   
\def\cite#1{\citealp{#1}}    

\def\authorindex#1{}

\begin{document}\newcount\preprintheader\preprintheader=1



\def\thisvolume{these proceedings}

\def\aj{{AJ}}			
\def\araa{{ARA\&A}}		
\def\apj{{ApJ}}			
\def\apjl{{ApJ}}		
\def\apjs{{ApJS}}		
\def\ao{{Appl.\ Optics}} 
\def\apss{{Ap\&SS}}		
\def\aap{{A\&A}}		
\def\aapr{{A\&A~Rev.}}		
\def\aaps{{A\&AS}}		
\def\an{{Astron.\ Nachrichten}}
\def\aspcs{{ASP Conf.\ Ser.}}
\def\azh{{AZh}}			
\def\baas{{BAAS}}		
\def\jrasc{{JRASC}}		
\def\memras{{MmRAS}}		
\def\mnras{{MNRAS}}
\def\nat{{Nat}}		
\def\pra{{Phys.\ Rev.\ A}} 
\def\prb{{Phys.\ Rev.\ B}}		
\def\prc{{Phys.\ Rev.\ C}}		
\def\prd{{Phys.\ Rev.\ D}}		
\def\prl{{Phys.\ Rev.\ Lett}}	
\def\pasp{{PASP}}
\def\pasj{{PASJ}}		
\def\qjras{{QJRAS}}
\def\science{{Sci}}		
\def\skytel{{S\&T}}		
\def\solphys{{Solar\ Phys.}} 
\def\sovast{{Soviet\ Ast.}}  
\def\ssr{{Space\ Sci.\ Rev.}}
\def\svassp{{Astrophys.\ Space Science Proc.}}
\def\zap{{ZAp}}			
\let\astap=\aap
\let\apjlett=\apjl
\let\apjsupp=\apjs

\def\ion#1#2{{\rm #1}\,{\uppercase{#2}}}  
\def\deg{\hbox{$^\circ$}}
\def\sun{\hbox{$\odot$}}
\def\earth{\hbox{$\oplus$}}
\def\la{\mathrel{\hbox{\rlap{\hbox{\lower4pt\hbox{$\sim$}}}\hbox{$<$}}}}
\def\ga{\mathrel{\hbox{\rlap{\hbox{\lower4pt\hbox{$\sim$}}}\hbox{$>$}}}}
\def\sq{\hbox{\rlap{$\sqcap$}$\sqcup$}}
\def\arcmin{\hbox{$^\prime$}}
\def\arcsec{\hbox{$^{\prime\prime}$}}
\def\fd{\hbox{$.\!\!^{\rm d}$}}
\def\fh{\hbox{$.\!\!^{\rm h}$}}
\def\fm{\hbox{$.\!\!^{\rm m}$}}
\def\fs{\hbox{$.\!\!^{\rm s}$}}
\def\fdg{\hbox{$.\!\!^\circ$}}
\def\farcm{\hbox{$.\mkern-4mu^\prime$}}
\def\farcs{\hbox{$.\!\!^{\prime\prime}$}}
\def\fp{\hbox{$.\!\!^{\scriptscriptstyle\rm p}$}}
\def\micron{\hbox{$\mu$m}}
\def\onehalf{\hbox{$\,^1\!/_2$}}	
\def\onethird{\hbox{$\,^1\!/_3$}}
\def\twothirds{\hbox{$\,^2\!/_3$}}
\def\onequarter{\hbox{$\,^1\!/_4$}}
\def\threequarters{\hbox{$\,^3\!/_4$}}
\def\ubv{\hbox{$U\!BV$}}		
\def\ubvr{\hbox{$U\!BV\!R$}}		
\def\ubvri{\hbox{$U\!BV\!RI$}}		
\def\ubvrij{\hbox{$U\!BV\!RI\!J$}}		
\def\ubvrijh{\hbox{$U\!BV\!RI\!J\!H$}}		
\def\ubvrijhk{\hbox{$U\!BV\!RI\!J\!H\!K$}}		
\def\ub{\hbox{$U\!-\!B$}}		
\def\bv{\hbox{$B\!-\!V$}}		
\def\vr{\hbox{$V\!-\!R$}}		
\def\ur{\hbox{$U\!-\!R$}}


\def\labelitemi{{\bf --}}  

\def\rmit#1{{\it #1}}              
\def\rmit#1{{\rm #1}}              
\def\etal{\rmit{et al.}}           
\def\etc{\rmit{etc.}}           
\def\ie{\rmit{i.e.,}}              
\def\eg{\rmit{e.g.,}}              
\def\cf{cf.}                       
\def\viz{\rmit{viz.}}
\def\vs{\rmit{vs.}}

\def\rot{\hbox{\rm rot}}
\def\div{\hbox{\rm div}}
\def\lesssim{\mathrel{\hbox{\rlap{\hbox{\lower4pt\hbox{$\sim$}}}\hbox{$<$}}}}
\def\gtrsim{\mathrel{\hbox{\rlap{\hbox{\lower4pt\hbox{$\sim$}}}\hbox{$>$}}}}
\def\dif{\: {\rm d}}                       
\def\ep{\:{\rm e}^}                        
\def\dash{\hbox{$\,-\,$}}                  
\def\is{\!=\!}                             

\def\starname#1#2{${#1}$\,{\rm {#2}}}  
\def\Teff{\hbox{$T_{\rm eff}$}}   

\def\kms{\hbox{km$\;$s$^{-1}$}}
\def\Mxcm{\hbox{Mx\,cm$^{-2}$}}    

\def\Bapp{\hbox{$B_{\rm app}$}}    

\def\komega{($k, \omega$)}                 
\def\kf{($k_h,f$)}                         
\def\VminI{\hbox{$V\!\!-\!\!I$}}           
\def\IminI{\hbox{$I\!\!-\!\!I$}}           
\def\VminV{\hbox{$V\!\!-\!\!V$}}           
\def\Xt{\hbox{$X\!\!-\!t$}}                

\def\level #1 #2#3#4{$#1 \: ^{#2} \mbox{#3} ^{#4}$}   

\def\specchar#1{\uppercase{#1}}    
\def\AlI{\mbox{Al\,\specchar{i}}}  
\def\BI{\mbox{B\,\specchar{i}}} 
\def\BII{\mbox{B\,\specchar{ii}}}  
\def\BaI{\mbox{Ba\,\specchar{i}}}  
\def\BaII{\mbox{Ba\,\specchar{ii}}} 
\def\CI{\mbox{C\,\specchar{i}}} 
\def\CII{\mbox{C\,\specchar{ii}}} 
\def\CIII{\mbox{C\,\specchar{iii}}} 
\def\CIV{\mbox{C\,\specchar{iv}}} 
\def\CaI{\mbox{Ca\,\specchar{i}}} 
\def\CaII{\mbox{Ca\,\specchar{ii}}} 
\def\CaIII{\mbox{Ca\,\specchar{iii}}} 
\def\CoI{\mbox{Co\,\specchar{i}}} 
\def\CrI{\mbox{Cr\,\specchar{i}}} 
\def\CriI{\mbox{Cr\,\specchar{ii}}} 
\def\CsI{\mbox{Cs\,\specchar{i}}} 
\def\CsII{\mbox{Cs\,\specchar{ii}}} 
\def\CuI{\mbox{Cu\,\specchar{i}}} 
\def\FeI{\mbox{Fe\,\specchar{i}}} 
\def\FeII{\mbox{Fe\,\specchar{ii}}} 
\def\FeIX{\mbox{Fe\,\specchar{ix}}}
\def\FeX{\mbox{Fe\,\specchar{x}}}
\def\FeXVI{\mbox{Fe\,\specchar{xvi}}}
\def\FrI{\mbox{Fr\,\specchar{i}}}
\def\HI{\mbox{H\,\specchar{i}}} 
\def\HII{\mbox{H\,\specchar{ii}}} 
\def\Hmin{\hbox{\rmH$^{^{_{\scriptstyle -}}}$}}      
\def\Hemin{\hbox{{\rm He}$^{^{_{\scriptstyle -}}}$}} 
\def\HeI{\mbox{He\,\specchar{i}}} 
\def\HeII{\mbox{He\,\specchar{ii}}} 
\def\HeIII{\mbox{He\,\specchar{iii}}} 
\def\KI{\mbox{K\,\specchar{i}}} 
\def\KII{\mbox{K\,\specchar{ii}}} 
\def\KIII{\mbox{K\,\specchar{iii}}} 
\def\LiI{\mbox{Li\,\specchar{i}}} 
\def\LiII{\mbox{Li\,\specchar{ii}}} 
\def\LiIII{\mbox{Li\,\specchar{iii}}} 
\def\MgI{\mbox{Mg\,\specchar{i}}} 
\def\MgII{\mbox{Mg\,\specchar{ii}}} 
\def\MgIII{\mbox{Mg\,\specchar{iii}}} 
\def\MnI{\mbox{Mn\,\specchar{i}}} 
\def\NI{\mbox{N\,\specchar{i}}}
\def\NaI{\mbox{Na\,\specchar{i}}}
\def\NaII{\mbox{Na\,\specchar{ii}}}
\def\NaIII{\mbox{Na\,\specchar{iii}}} 
\def\NiI{\mbox{Ni\,\specchar{i}}} 
\def\NiII{\mbox{Ni\,\specchar{ii}}}
\def\NiIII{\mbox{Ni\,\specchar{iii}}} 
\def\OI{\mbox{O\,\specchar{i}}} 
\def\OVI{\mbox{O\,\specchar{vi}}}
\def\RbI{\mbox{Rb\,\specchar{i}}} 
\def\SII{\mbox{S\,\specchar{ii}}} 
\def\SiI{\mbox{Si\,\specchar{i}}} 
\def\SiII{\mbox{Si\,\specchar{ii}}} 
\def\SrI{\mbox{Sr\,\specchar{i}}}
\def\SrII{\mbox{Sr\,\specchar{ii}}}
\def\TiI{\mbox{Ti\,\specchar{i}}} 
\def\TiII{\mbox{Ti\,\specchar{ii}}} 
\def\TiIII{\mbox{Ti\,\specchar{iii}}} 
\def\TiIV{\mbox{Ti\,\specchar{iv}}} 
\def\VI{\mbox{V\,\specchar{i}}} 
\def\HtwoO{\mbox{H$_2$O}}        
\def\Otwo{\mbox{O$_2$}}          

\def\Halpha{\mbox{H\hspace{0.1ex}$\alpha$}} 
\def\Ha{\mbox{H\hspace{0.2ex}$\alpha$}}
\def\Hbeta{\mbox{H\hspace{0.2ex}$\beta$}}
\def\Hgamma{\mbox{H\hspace{0.2ex}$\gamma$}}
\def\Hdelta{\mbox{H\hspace{0.2ex}$\delta$}}
\def\Hepsilon{\mbox{H\hspace{0.2ex}$\epsilon$}}
\def\Hzeta{\mbox{H\hspace{0.2ex}$\zeta$}}
\def\Lyalpha{\mbox{Ly$\hspace{0.2ex}\alpha$}}
\def\Lybeta{\mbox{Ly$\hspace{0.2ex}\beta$}}
\def\Lygamma{\mbox{Ly$\hspace{0.2ex}\gamma$}}
\def\Lycont{\mbox{Ly\hspace{0.2ex}{\small cont}}}
\def\Baalpha{\mbox{Ba$\hspace{0.2ex}\alpha$}}
\def\Babeta{\mbox{Ba$\hspace{0.2ex}\beta$}}
\def\Bacont{\mbox{Ba\hspace{0.2ex}{\small cont}}}
\def\Paalpha{\mbox{Pa$\hspace{0.2ex}\alpha$}}
\def\Bralpha{\mbox{Br$\hspace{0.2ex}\alpha$}}

\def\NaD{\mbox{Na\,\specchar{i}\,D}}    
\def\NaDone{\mbox{Na\,\specchar{i}\,\,D$_1$}}
\def\NaDtwo{\mbox{Na\,\specchar{i}\,\,D$_2$}}
\def\NaID{\mbox{Na\,\specchar{i}\,\,D}}
\def\NaIDone{\mbox{Na\,\specchar{i}\,\,D$_1$}}
\def\NaIDtwo{\mbox{Na\,\specchar{i}\,\,D$_2$}}
\def\Done{\mbox{D$_1$}}
\def\Dtwo{\mbox{D$_2$}}

\def\Mgbone{\mbox{Mg\,\specchar{i}\,b$_1$}}
\def\Mgbtwo{\mbox{Mg\,\specchar{i}\,b$_2$}}
\def\Mgbthree{\mbox{Mg\,\specchar{i}\,b$_3$}}
\def\MgIb{\mbox{Mg\,\specchar{i}\,b}}
\def\MgIbone{\mbox{Mg\,\specchar{i}\,b$_1$}}
\def\MgIbtwo{\mbox{Mg\,\specchar{i}\,b$_2$}}
\def\MgIbthree{\mbox{Mg\,\specchar{i}\,b$_3$}}

\def\CaIIK{\mbox{Ca\,\specchar{ii}\,K}}       
\def\CaIIH{\mbox{Ca\,\specchar{ii}\,H}}
\def\CaIIHK{\mbox{Ca\,\specchar{ii}\,H\,\&\,K}}
\def\HK{\mbox{H\,\&\,K}}
\def\Kthree{\mbox{K$_3$}}      
\def\Hthree{\mbox{H$_3$}}
\def\Ktwo{\mbox{K$_2$}}
\def\Htwo{\mbox{H$_2$}}
\def\Kone{\mbox{K$_1$}}     
\def\Hone{\mbox{H$_1$}}     
\def\KtwoV{\mbox{K$_{2V}$}}
\def\KtwoR{\mbox{K$_{2R}$}}
\def\KoneV{\mbox{K$_{1V}$}}
\def\KoneR{\mbox{K$_{1R}$}}
\def\HtwoV{\mbox{H$_{2V}$}}
\def\HtwoR{\mbox{H$_{2R}$}}
\def\HoneV{\mbox{H$_{1V}$}}
\def\HoneR{\mbox{H$_{1R}$}}

\def\hk{\mbox{h\,\&\,k}}
\def\kthree{\mbox{k$_3$}}    
\def\hthree{\mbox{h$_3$}}
\def\ktwo{\mbox{k$_2$}}
\def\htwo{\mbox{h$_2$}}
\def\kone{\mbox{k$_1$}}     
\def\hone{\mbox{h$_1$}}     
\def\ktwoV{\mbox{k$_{2V}$}}
\def\ktwoR{\mbox{k$_{2R}$}}
\def\koneV{\mbox{k$_{1V}$}}
\def\koneR{\mbox{k$_{1R}$}}
\def\htwoV{\mbox{h$_{2V}$}}
\def\htwoR{\mbox{h$_{2R}$}}
\def\honeV{\mbox{h$_{1V}$}}
\def\honeR{\mbox{h$_{1R}$}}

\title*{Major Surge Activity of Super-Active Region NOAA 10484}

\titlerunning{Surge Activity from NOAA 10484}  

\author{W. Uddin\inst{1}
        \and
         P. Kumar\inst{1}
        \and
        A. K. Srivastava\inst{1}
	\and
	 R. Chandra\inst{2}}

\authorindex{Uddin, W.}
\authorindex{Kumar, P.}
\authorindex{Srivastava, A. K.}
\authorindex{Chandra, R.}

\authorrunning{Uddin et al.}  

\institute{Aryabhatta Research Institute of Observational Sciences, 
           Nainital, India.
           \and 
           Observatoire de Paris, LESIA, Meudon, France.}

\maketitle

\setcounter{footnote}{0}  

\begin{abstract} 
  We observed two surges in H$\alpha$ from the super-active region
  NOAA 10484. The first surge was associated with an SF/C4.3 class
  flare. The second one was a major surge associated with a SF/C3.9
  flare. This surge was also observed with SOHO/EIT in 195~\AA\ and
  NoRh in 17~GHz, and showed similar evolution in these wavelengths.
  The major surge had an ejective funnel-shaped spray structure with
  fast expansion in linear (about $1.2 \times 10^5$~km) and angular
  (about 65\deg) size during its maximum phase. The mass motion of the
  surge was along open magnetic field lines, with average velocity
  about 100~\kms.  The de-twisting motion of the surge reveals
  relaxation of sheared and twisted magnetic flux.  The SOHO/MDI
  magnetograms reveal that the surges occurred at the site of
  companion sunspots where positive flux emerged, converged, and
  canceled against surrounding field of opposite polarity.  Our
  observations support magnetic reconnection models for the surges and
  jets.
\end{abstract}

\section{Introduction}      \label{uddin-sec:introduction}
During October--November 2003, major solar activity originated from
three super-active regions, namely NOAA AR 10484, 10486 and 10488. On
October 25, we observed two surges between 01:50 UT and 04:15 UT which
originated from NOAA AR 10484. The first surge was small; the second
one was very dynamic and explosive in nature.  Using multi-wavelength
data we present a morphological study of these surges in order to
understand the physical processes behind their activity.

\begin{figure}
\centering
\includegraphics[width=11.8 cm]{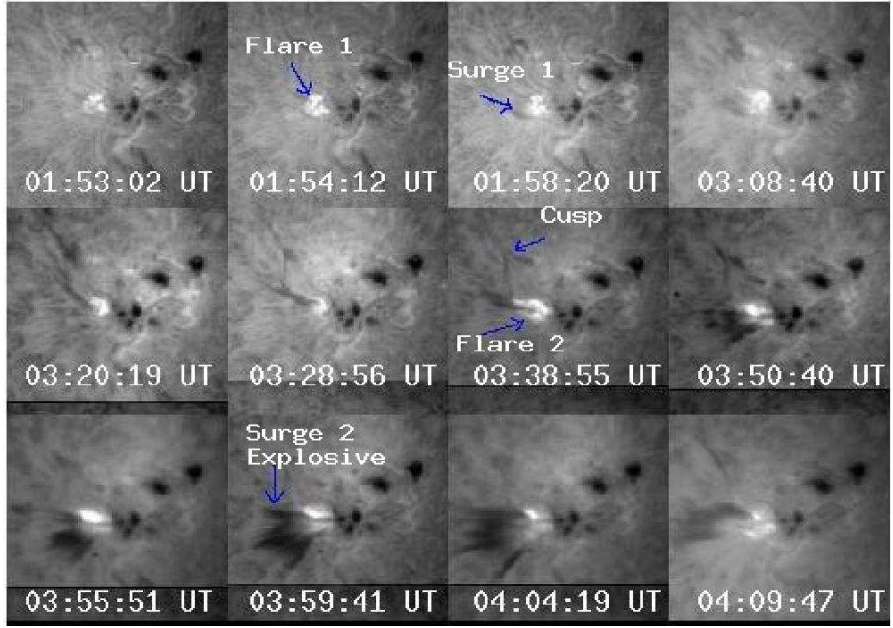}
\caption[]{\label{uddin-fig1}
  Sample images of surge evolution in H$\alpha$.  The field of view is
  is $300\arcsec\times300\arcsec$.
}\end{figure}

\section{Observations}                   \label{uddin-sec:Observations}


H$\alpha$ images of these events were obtained at Aryabhatta Research
Institute of Observational Sciences (ARIES), Nainital, India on 25
October 2003 using the 15-cm f/15 Coud\'e solar tower telescope
equipped with a Bernard Halle H$\alpha$ filter, at intervals of
15-20~s and with a pixel size of 1\arcsec.  We also used data from
SOHO/MDI (cadence 96~min, 1.98\arcsec\ pixels, \cite{Scherrer95}),
SOHO/EIT (cadence 12~min, 2.5\arcsec\ pixels, \cite{Delaboudiniere95})
and NoRh (cadence 10~s, 5\arcsec\ pixels, \cite{Takano97}).

\begin{figure}
\centering
\includegraphics[width=10cm]{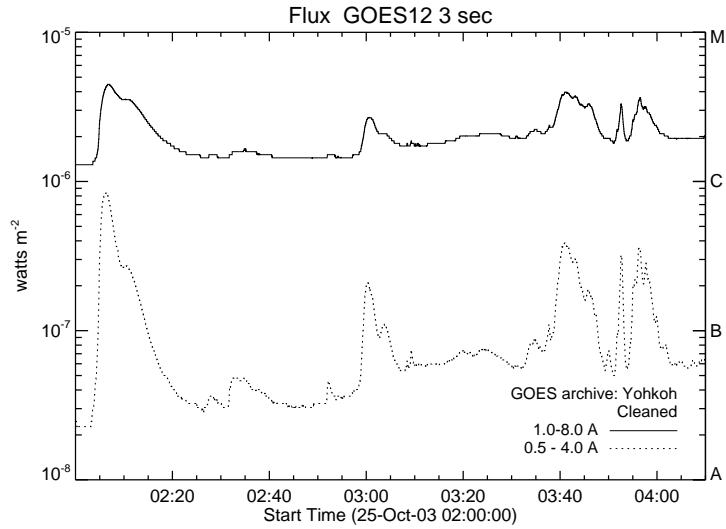}
\vspace*{-0.3cm}
\caption[]{\label{uddin-fig2}
  GOES time profiles of the 25 October 2003 flares from NOAA AR 10484.
  }\end{figure}

The H$\alpha$ observations nicely show the dynamic evolution of the
recurrent surge activities from O1:50~UT to 04:15~UT
(Fig.~\ref{uddin-fig1}). The surge activities occurred at the
following satellite sunspot of the active region. First, a small surge
was associated with a small (SF/C4.3) flare which started at 01:55~UT,
reached maximum at 01:57~UT, and ended at 01:59~UT). The arrows show
flare 1 and surge 1 in the H$\alpha$ images. Another subflare
(SF/C2.6) then started at 02:59~UT, peaked at 03:00~UT, and ended at
03:07~UT without surge activity.  At 03:32~UT, another eruptive
subflare (SF/C3.9) started with the second, major, dynamic and
explosive surge.  It reached maximum at 03:52~UT and continued up to
04:15~UT. The soft X-ray flux showed two flares during this main surge
eruption. The temporal evolution of these 25 October, 2003 flares from
NOAA AR 10484 is presented in Fig.~\ref{uddin-fig2}. The surge evolved
with initially small velocity, but in the ascending phase its velocity
grew and it showed funnel-like structures during its maximum phase at
03:59~UT, which indicate spray-type behaviour (the arrow indicates
this explosive surge 2 in Fig.~\ref{uddin-fig1}). The two-ribbon
structure at the footpoint of the surge was also seen during the surge
eruption. The H$\alpha$ movie shows the change in orientation of the
surge (from North-East to South-West) and also de-twisting motion was
observed during its evolution.

\begin{figure}
  \centering \includegraphics[width=2.85cm]{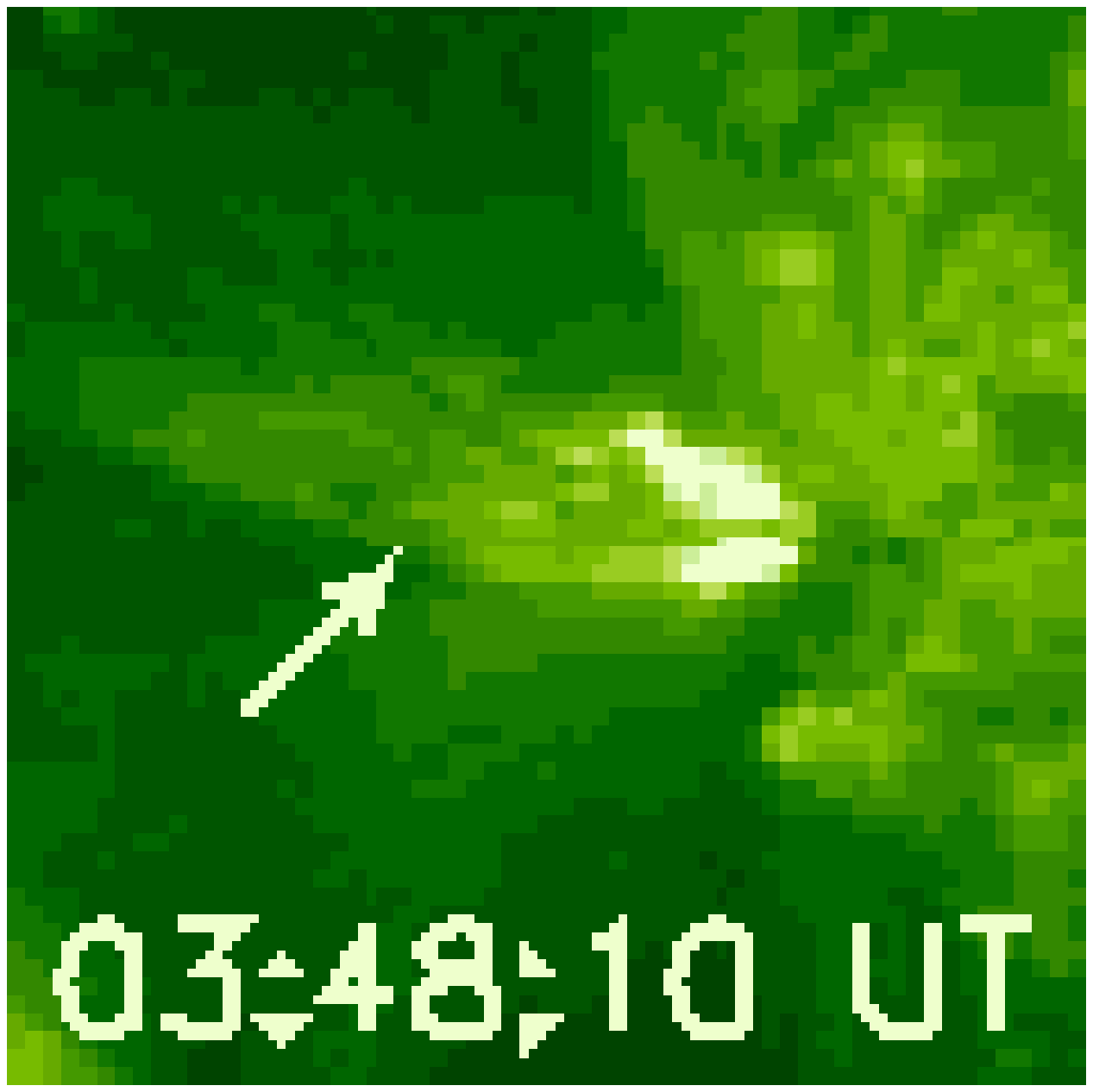}
  \includegraphics[width=2.85cm]{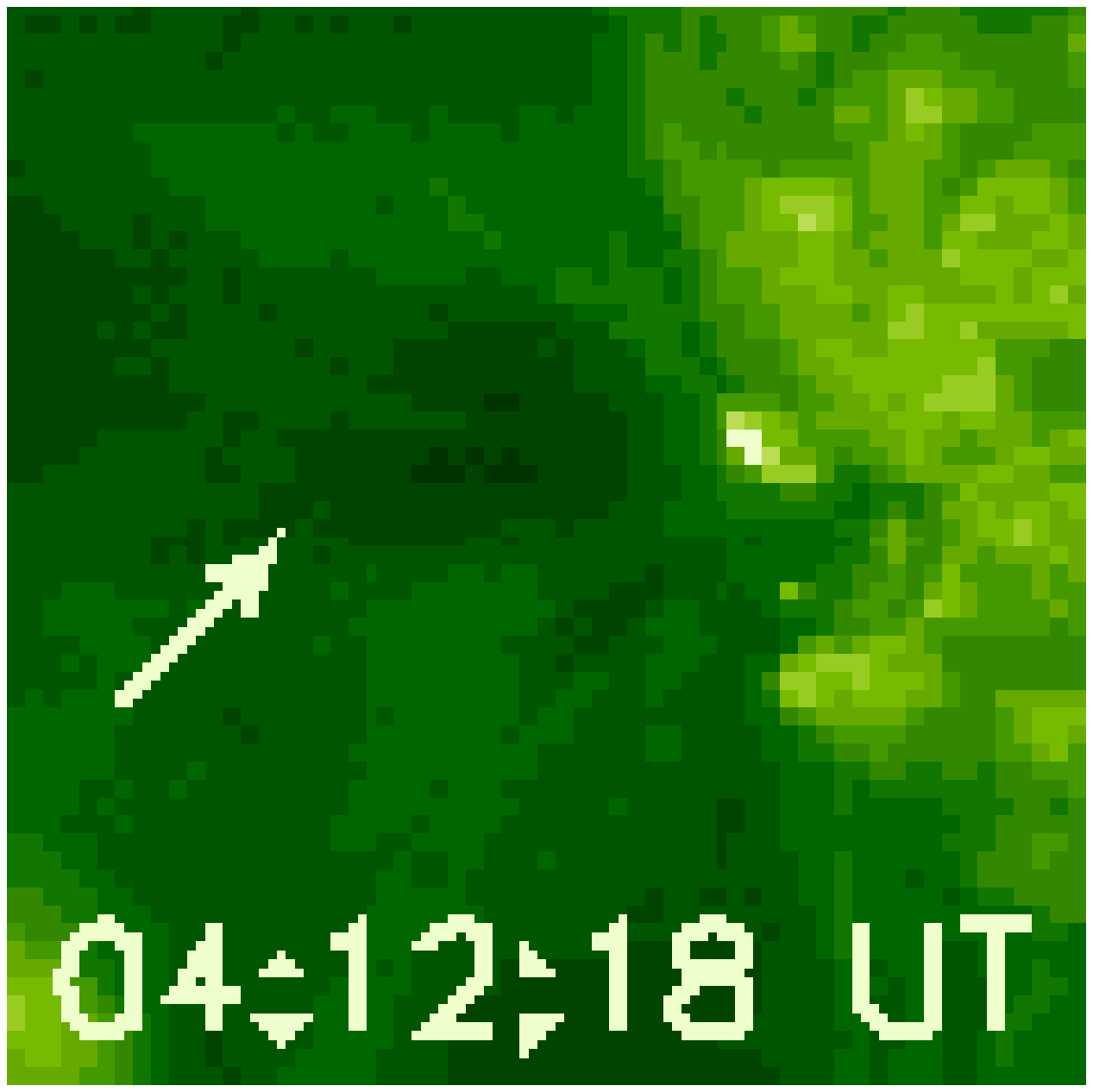}
  \includegraphics[width=2.85cm]{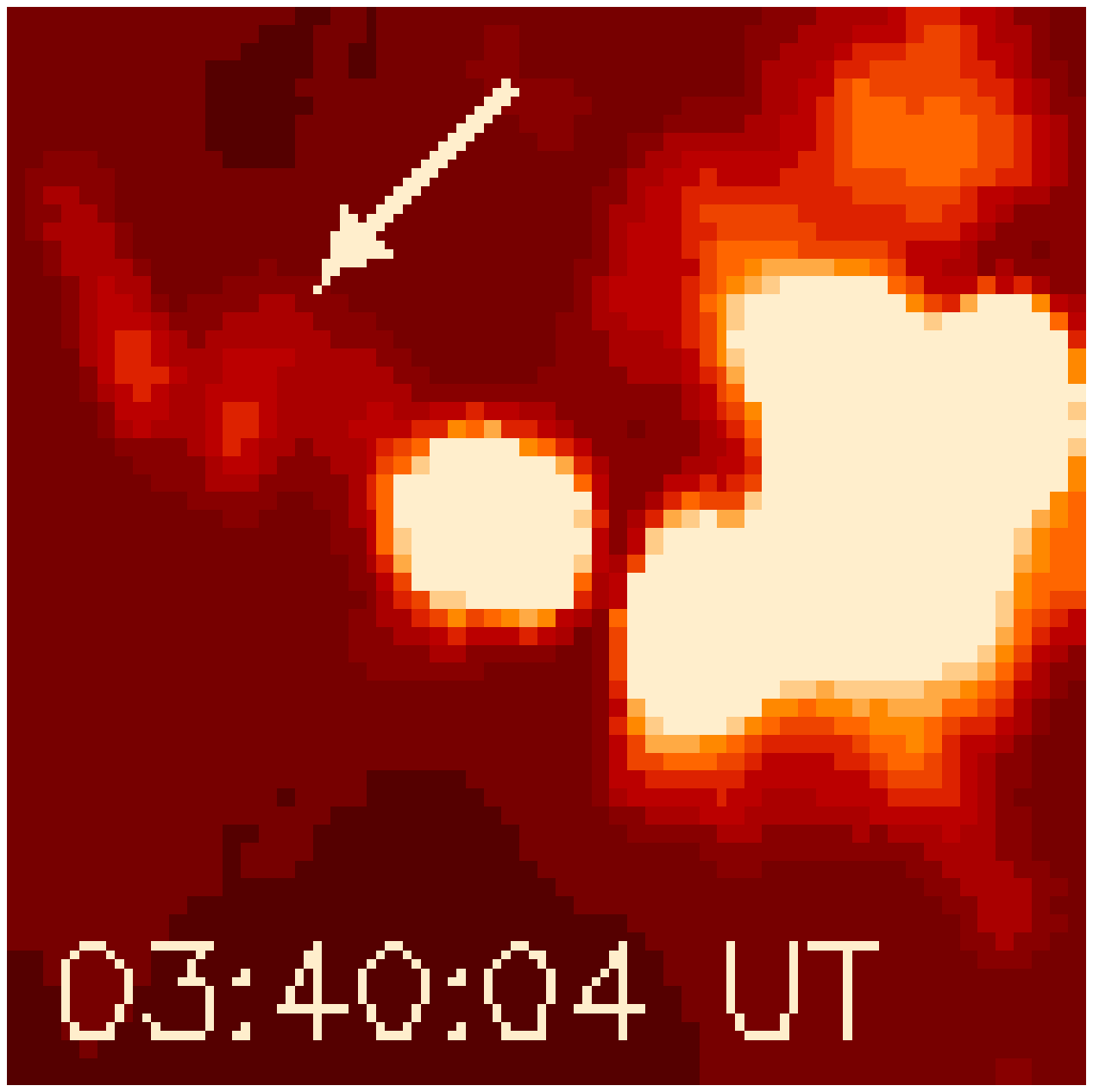}
  \includegraphics[width=2.85cm]{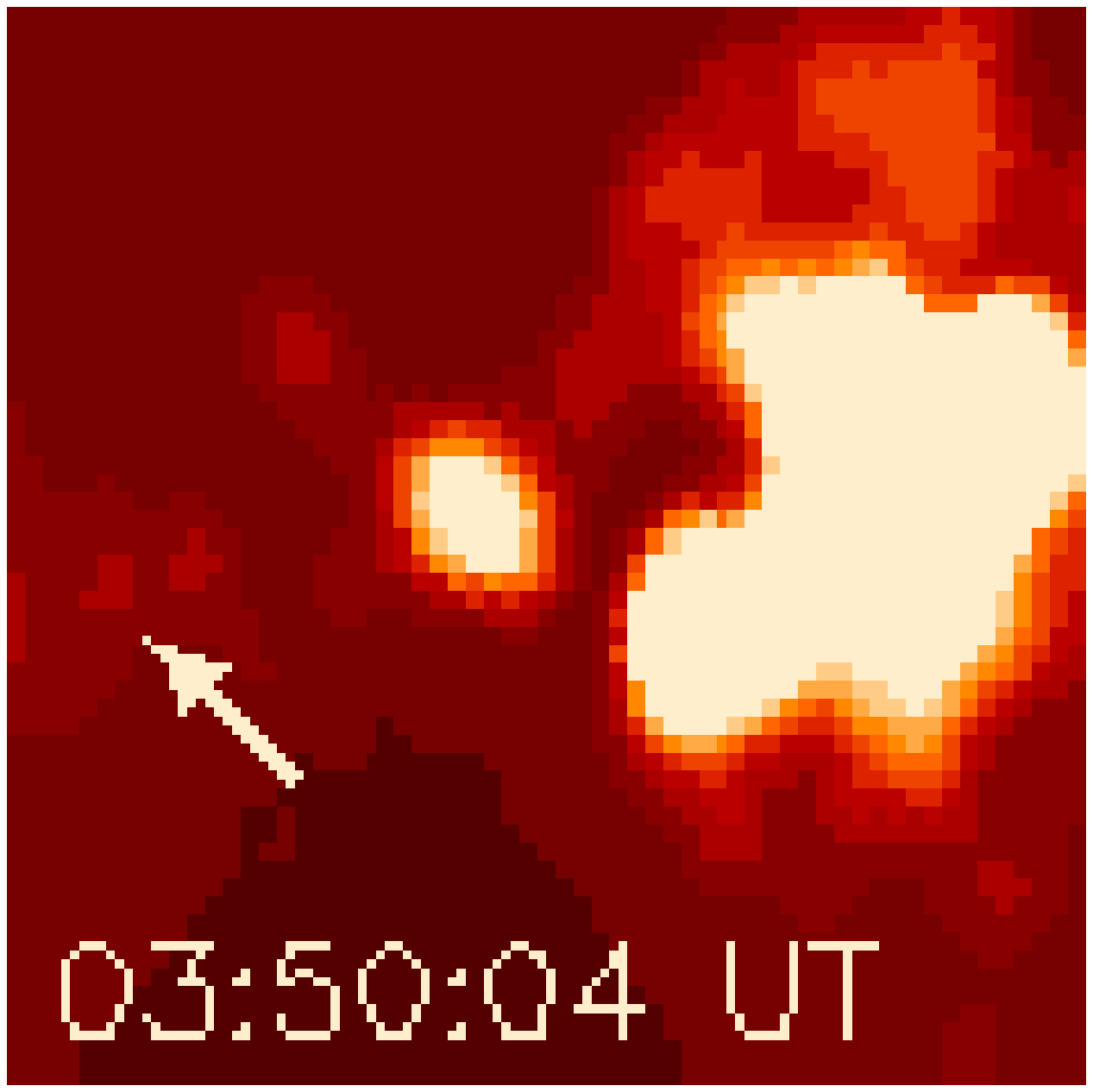}
  \includegraphics[width=2.85cm]{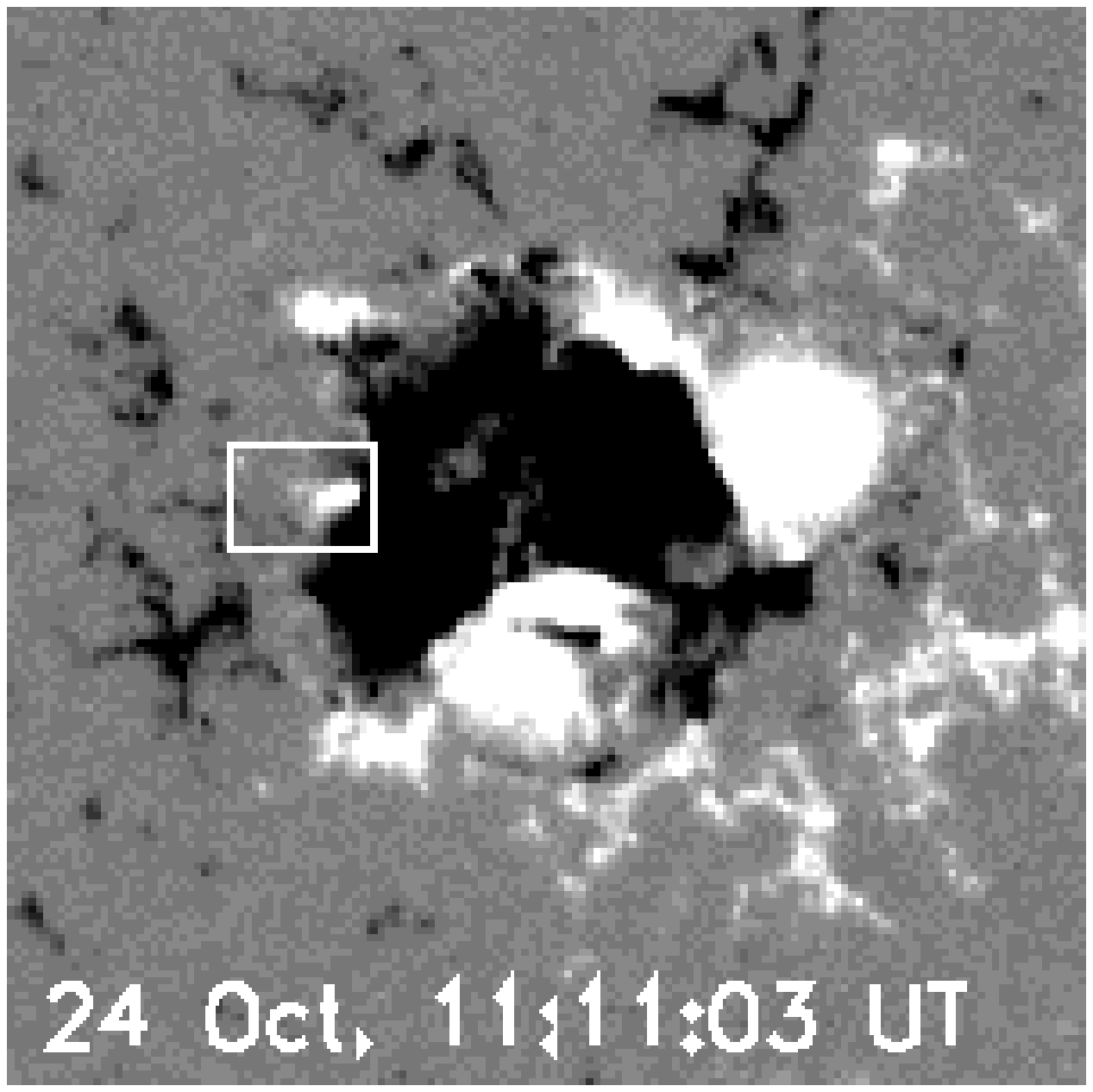}
  \includegraphics[width=2.85cm]{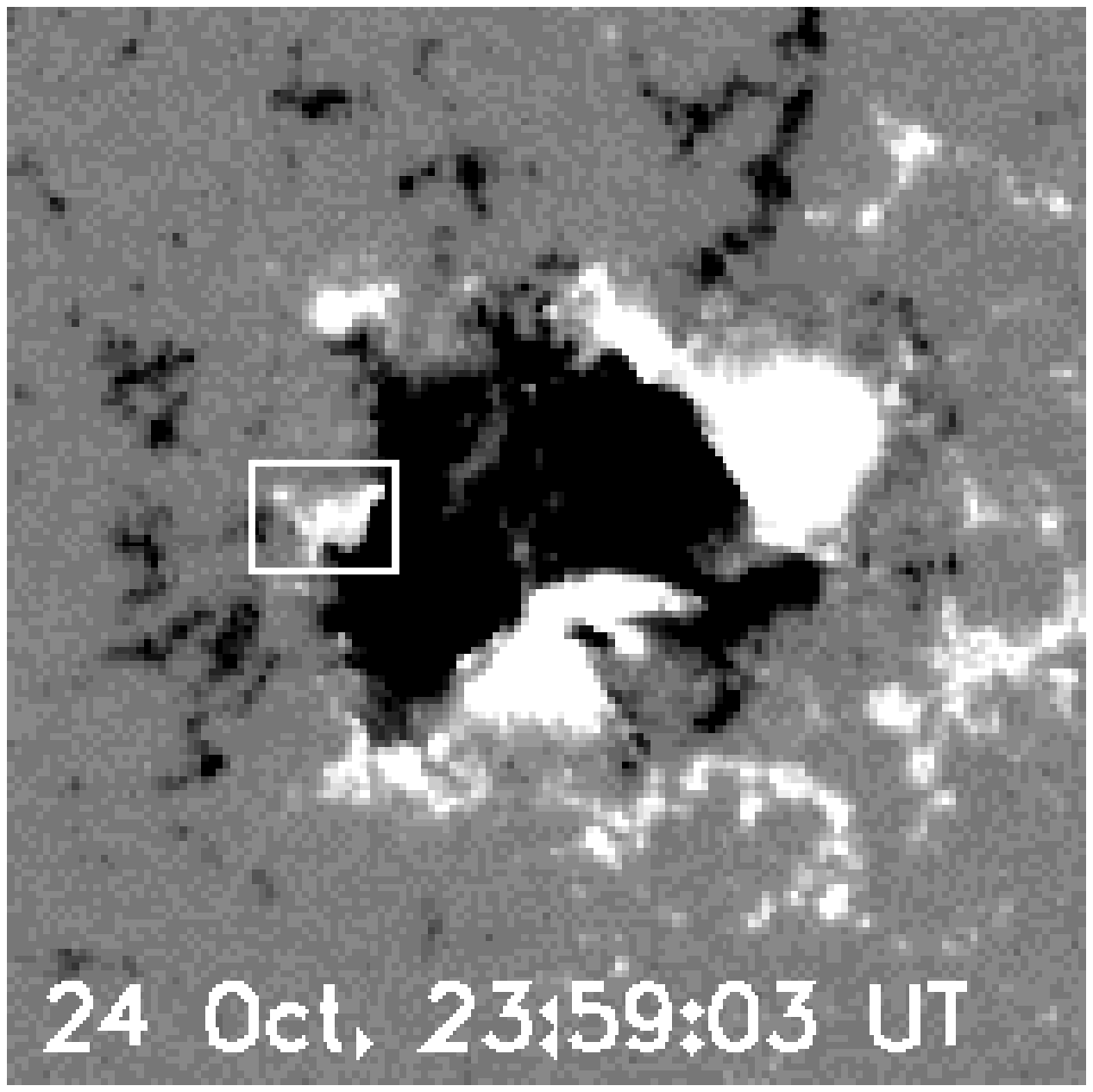}
  \includegraphics[width=2.85cm]{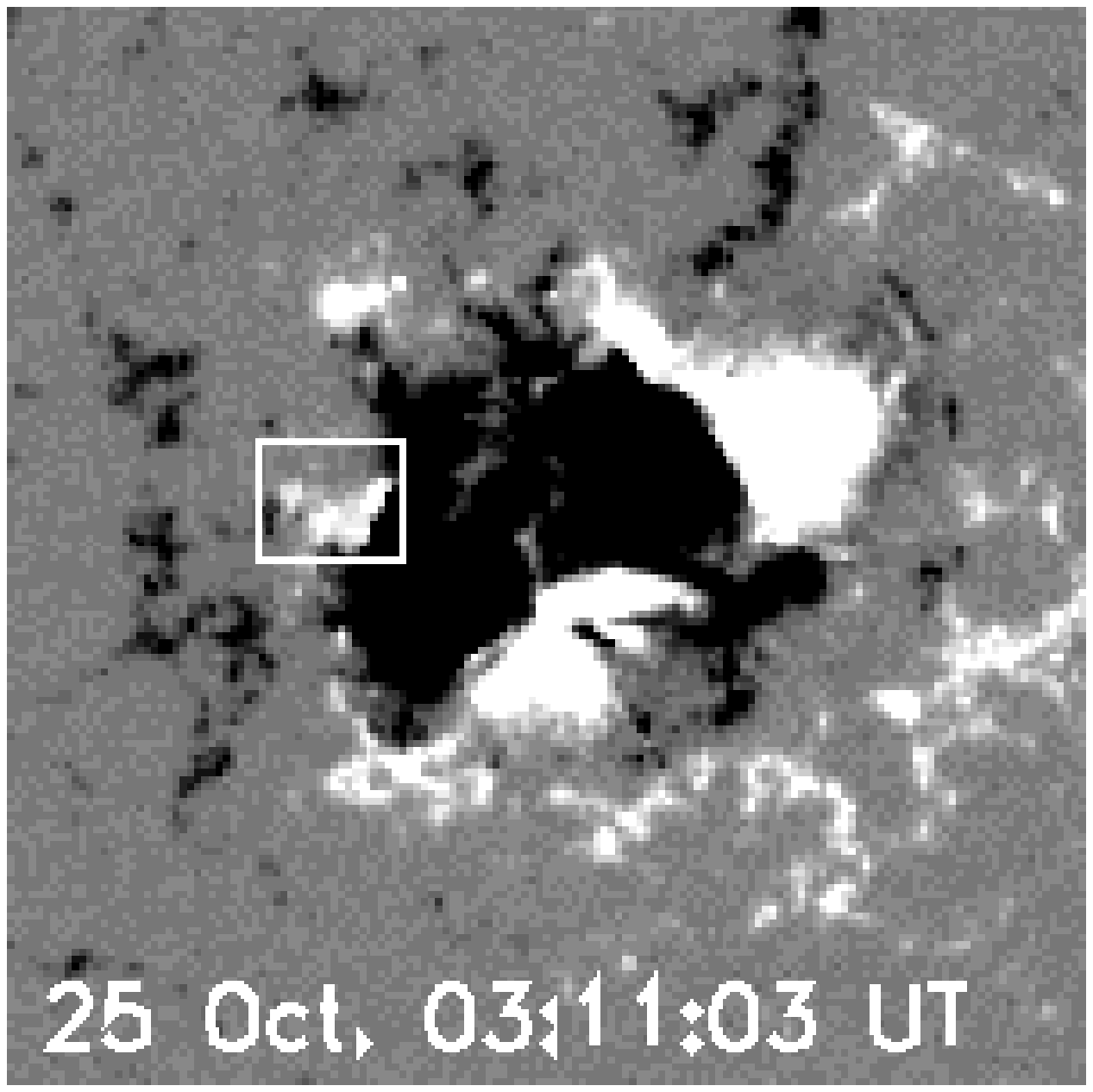}
  \includegraphics[width=2.85cm]{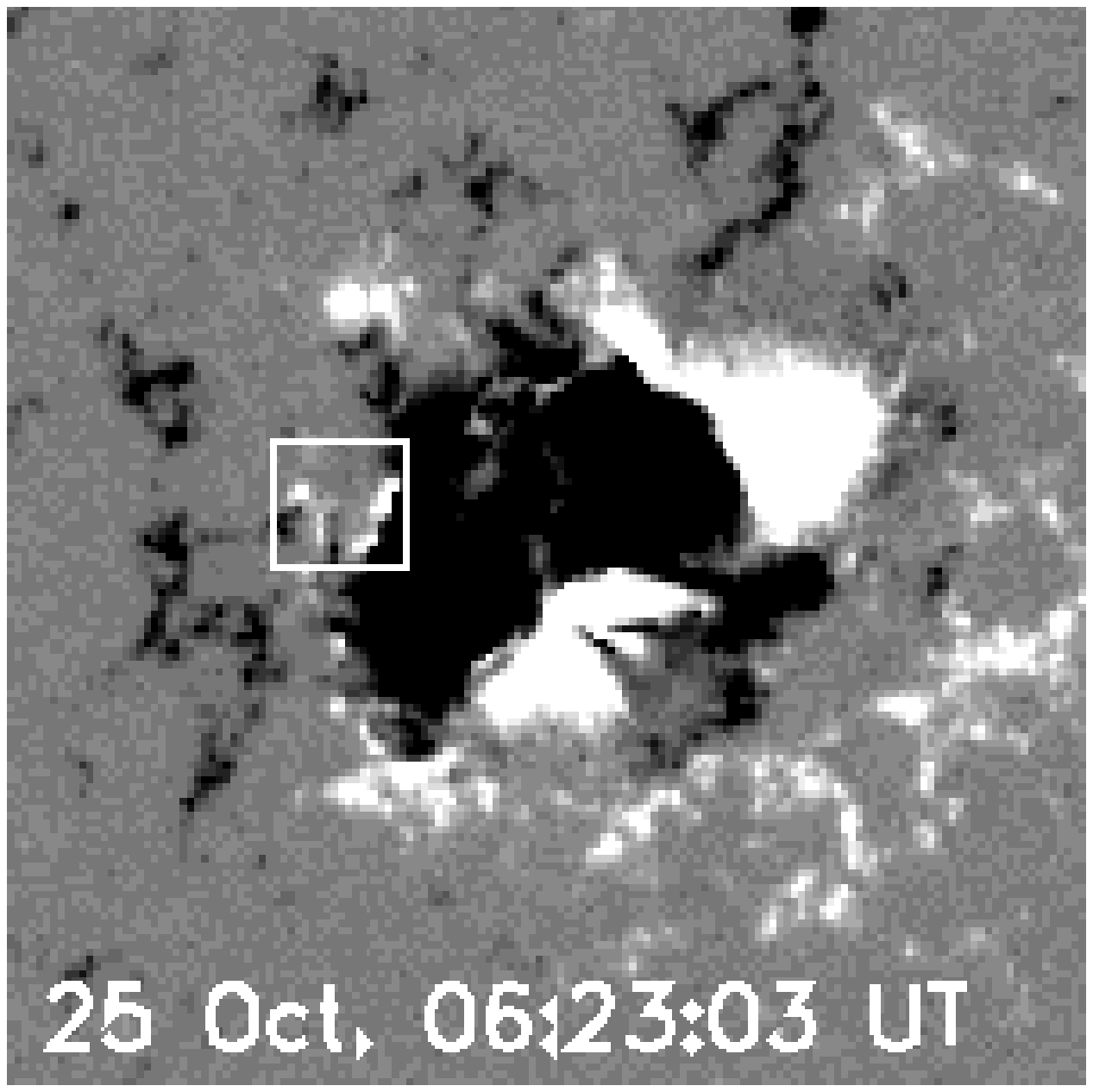}
  \caption[]{\label{uddin-fig3} Top: two EIT 195~\AA\ images and
  and two NoRh 17~GHz images showing the surge eruption.
  Bottom: SOHO/MDI magnetograms which show flux emergence and
  cancellation within the box.  The field of view of
  each frame is $300\arcsec\times300\arcsec$.  }
\end{figure}

The EIT 195~\AA observations show similar morphology of the surge as
in H$\alpha$: two-ribbon structure and mass motion at 03:48~UT (arrow
in Fig.~\ref{uddin-fig3}). The Nobeyama 17~GHz images also show the
orientation change of the surge material as being similar to that in
the H$\alpha$ images at 03:40~UT (Fig.~\ref{uddin-fig3}).

The MDI observations show positive flux emergence before the 5--6~hours
of surge activity nearby the satellite sunspot (marked by the box
in Fig.~\ref{uddin-fig3}). This flux emerged before the surge
activity and disappeared after the event.

\section{Results and discussion}                   \label{uddin-Results and Discussion}

From our detailed investigation it is evident that the surges were
associated with many C-class subflares, which indicates magnetic field
annihilation at the site of surge activity. The MDI magnetograms
reveal positive flux emergence and its cancellation by surrounding
opposite polarity fields. The major surge rose upwards with an average
velocity of about 100~\kms; its orientation change and de-twisting
motion demonstrated relaxation of sheared and twisted magnetic field.
The funnel-shaped structure of the surge is due to the material that
follows the open magnetic field lines at the site (visible in the EIT
images). The surge shows similar evolution in H$\alpha$ (chromospheric
temperature), EIT 195~\AA\ (coronal temperature), and Nobeyama radio
observations at 17~GHz (non-thermal cornal emission).

These multi-wavelength data indicate that the first reconnection took
place between the newly emerged positive polarity sunspot and the
pre-existing surrounding field.  Subflaring then occurred, plasma was
heated up to 10~MK, and transported towards open and closed field
lines, which led to the formation of the two small flare loops visible in
H$\alpha$ and EIT, and the funnel-shaped surge structures.
This scenario was earlier reported by
\citet{uddin-1995Nature.375...42Y} on the basis of numerical
simulations.  The flaring loops that formed nearby the footpoint of
the surge and the type-III radio burst that was observed during this
event are evidence favoring the magnetic-reconnection surge model 
(\cite{uddin-1994ApJ...431L..51S}, \cite{uddin-1995SoPh..156..245S},
\cite{uddin-1996ApJ...464.1016C}).  Our observational results support
this model of magnetic reconnection for surges. These are only the
preliminary results; we are planning to carry out more detailed
study of these observations.

\begin{acknowledgement}
  We thank the conference organisers for a very good meeting and the
  editors for excellent instructions. R.C. thanks the CIFIPRA for his
  postdoc grant.
\end{acknowledgement}


\begin{small}



\begin{thebibliography}{7}
\expandafter\ifx\csname natexlab\endcsname\relax\def\natexlab#1{#1}\fi

\bibitem[{{Canfield} {et~al.}(1996){Canfield}, {Reardon}, {Leka}, {Shibata},
  {Yokoyama}, \& {Shimojo}}]{uddin-1996ApJ...464.1016C} {Canfield},
  R.~C., {Reardon}, K.~P., {Leka}, K.~D., {et~al.} 1996, \apj, 464,
  1016

\bibitem[{{Delaboudini{\`e}re} {et~al.}(1995){Delaboudini{\`e}re}, {Artzner},
  {Brunaud}, {Gabriel}, {Hochedez}, {Millier}, {Song}, {Au}, {Dere},
  {Howard}, {Kreplin}, {Michels}, {Moses}, {Defise}, {Jamar},
  {Rochus}, {Chauvineau}, {Marioge}, {Catura}, {Lemen}, {Shing},
  {Stern}, {Gurman}, {Neupert}, {Maucherat}, {Clette}, {Cugnon}, \&
  {van Dessel}}]{Delaboudiniere95} {Delaboudini{\`e}re}, J.-P.,
  {Artzner}, G.~E., {Brunaud}, J., {et~al.} 1995, \solphys, 162, 291

\bibitem[{{Scherrer} {et~al.}(1995){Scherrer}, {Bogart}, {Bush}, {Hoeksema},
  {Kosovichev}, {Schou}, {Rosenberg}, {Springer}, {Tarbell}, {Title},
  {Wolfson}, {Zayer}, \& {MDI Engineering Team}}]{Scherrer95}
  {Scherrer}, P.~H., {Bogart}, R.~S., {Bush}, R.~I., {et~al.} 1995,
  \solphys, 162, 129

\bibitem[{{Schmieder} {et~al.}(1995){Schmieder}, {Shibata}, {van
  Driel-Gesztelyi}, \& {Freeland}}]{uddin-1995SoPh..156..245S}
  {Schmieder}, B., {Shibata}, K., {van Driel-Gesztelyi}, L.,
  {Freeland}, S. 1995, \solphys, 156, 245

\bibitem[{{Shibata} {et~al.}(1994){Shibata}, {Nitta}, {Strong}, {Matsumoto},
  {Yokoyama}, {Hirayama}, {Hudson}, \&
  {Ogawara}}]{uddin-1994ApJ...431L..51S} {Shibata}, K., {Nitta}, N.,
  {Strong}, K.~T., {et~al.} 1994, \apjl, 431, L51

\bibitem[{{Takano} {et~al.}(1997){Takano}, {Nakajima}, {Enome}, {Shibasaki},
  {Nishio}, {Hanaoka}, {Shiomi}, {Sekiguchi}, {Kawashima},
  {Bushimata}, {Shinohara}, {Torii}, {Fujiki}, \&
  {Irimajiri}}]{Takano97} {Takano}, T., {Nakajima}, H., {Enome}, S.,
  {et~al.} 1997, in Coronal Physics from Radio and Space Observations,
  ed. G.~{Trottet}, Lecture Notes in Physics, 483, 183

\bibitem[{{Yokoyama} \& {Shibata}(1995)}]{uddin-1995Nature.375...42Y}
{Yokoyama}, T. {Shibata}, K. 1995, \nat, 375, 42

\end{thebibliography}

\end{small}
\end{document}